\documentclass[twocolumn,amssymb, amsmath, aps, superscriptaddress, showpacs, footinbib,  prb]{revtex4}
\usepackage{graphicx}

\newcommand{\CW}{\text{CW}}
\newcommand{\mg}{\text{mag}}

\begin{document}

\title{Magnetic properties of BaCdVO(PO$_4$)$_2$: a strongly frustrated\\ spin-1/2 square lattice close to the quantum critical regime}
\author{R. Nath}
\email{ramesh{_}phy2003@yahoo.com}
\affiliation{Max Planck Institute for Chemical Physics of Solids, N\"{o}thnitzer Str. 40, 01187 Dresden, Germany}
\author{A. A. Tsirlin}
\email{altsirlin@gmail.com}
\affiliation{Max Planck Institute for Chemical Physics of Solids, N\"{o}thnitzer
Str. 40, 01187 Dresden, Germany}
\affiliation{Department of Chemistry, Moscow State University, 119992 Moscow, Russia}
\author{H. Rosner}
\author{C. Geibel}
\affiliation{Max Planck Institute for Chemical Physics of Solids, N\"{o}thnitzer
Str. 40, 01187 Dresden, Germany}

\begin{abstract}
We report magnetization and specific heat measurements on polycrystalline
samples of BaCdVO(PO$_{4})_{2}$ and show that this compound is a $S=1/2$
frustrated square lattice with ferromagnetic nearest-neighbor ($J_{1}$) and
antiferromagnetic next-nearest-neighbor ($J_{2}$) interactions. The coupling
constants $J_{1}\simeq -3.6$~K and $J_{2}\simeq 3.2$~K are determined from a
fitting of the susceptibility data and confirmed by an analysis of the
saturation field ($\mu_0H_{s}=4.2$ T), the specific heat, and the magnetic
entropy. BaCdVO(PO$_{4})_{2}$ undergoes magnetic ordering at about 1 K,
likely towards a columnar antiferromagnetic state. We find that BaCdVO(PO$%
_{4})_{2}$ with the frustration ratio $\alpha =J_{2}/J_{1}\simeq -0.9$ is
closer to a critical (quantum spin liquid) region of the frustrated square
lattice than any of the previously reported compounds. Positive curvature of
the magnetization curve is observed in agreement with recent theoretical
predictions for high-field properties of the frustrated square lattice close
to the critical regime.
\end{abstract}

\maketitle

\section{Introduction}

\label{intro} Low-dimensional spin systems are one of the actively studied subjects in solid state physics due to the possibility to observe numerous quantum phenomena and to interpret these phenomena within relatively simple models (e.g., Ising or Heisenberg models for different lattice types). An interesting phenomenon in the spin physics is the formation of a spin liquid -- a strongly correlated ground state lacking long-range magnetic order. This ground state is usually related to the electronic mechanism of superconductivity suggested for high-$T_{c}$ cuprates.\cite{anderson1987,lee2008} Spin liquids originate from quantum fluctuations that are particularly strong in systems with reduced dimensionality and low spin value. The fluctuations can be further enhanced by introducing magnetic frustration which impedes long-range ordering of the system.

\begin{figure}[b]
\includegraphics{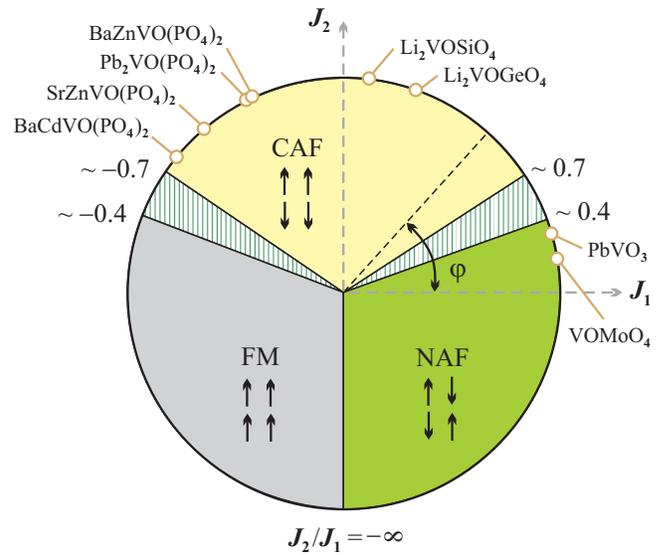}
\caption{\label{diagram1}
(Color online) Phase diagram of the FSL model.\cite{schmidt2007a} Solid filling indicates the regions of long-range magnetic ordering, while hatched filling denotes the critical regions. Positions of BaCdVO(PO$_4)_2$ and some of the previously investigated compounds are shown (see text for references).}
\end{figure}

The spin-1/2 frustrated square lattice (FSL) is one of the simplest models
giving rise to a spin liquid ground state. In this model (also known as
the $J_{1}-J_{2}$ model), magnetic moments on a square-lattice are subjected
to nearest-neighbor interaction $J_{1}$ along the side of the square
and next-nearest-neighbor interaction $J_{2}$ along the diagonal of
the square. The FSL model is described by the frustration ratio $\alpha
=J_{2}/J_{1}$ or, alternatively, by the frustration angle $\varphi =$$\text{%
tan}$$^{-1}(J_{2}/J_{1})$. Defining the thermodynamic energy scale of exchange couplings as $%
J_{c}=\sqrt{J_{1}^{2}+J_{2}^{2}}$, one obtains $J_{1}=J_{c}\cos
\varphi $ and $J_{2}=J_{c}\sin \varphi $ (see Fig.~\ref{diagram1} and Ref.~\onlinecite{shannon2004}).

Extensive theoretical research on the FSL model has been done in the past. Initially, the
studies were focused on the AFM region ($J_{1},J_{2}>0$),\cite%
{chandra1988,bacci1991,sushkov2001,siurakshina2001} while the general case
(arbitrary signs for $J_{1}$ and $J_{2}$) was only considered quite recently.%
\cite{shannon2004,shannon2006,schmidt2007a,schmidt2007b,schmidt2008} The
phase diagram of the model (Fig.~\ref{diagram1}) includes three regions with
different ordered phases: ferromagnet [FM, wave vector $\mathbf{Q}=(0,0)$], N\'{e}el antiferromagnet [NAF, $\mathbf{Q}=(\pi,\pi)$], and columnar antiferromagnet\cite{foot3} [CAF, $\mathbf{Q}=(\pi ,0)$ or $(0,\pi)$]. Classically, first-order phase transitions should occur at the NAF--CAF ($\alpha =0.5$) and the CAF--FM ($\alpha =-0.5$) boundaries. However, quantum fluctuations
destroy long-range ordering, leading to the formation of critical regions with disordered ground states. In general, the ground state in the critical regions is referred to a quantum spin liquid (QSL) regime, but the particular nature of the spin liquid phases is still under discussion. A gapless nematic state is suggested for $\alpha \sim -0.5$,\cite{shannon2006} while different
dimer phases (including resonating-valence-bond-type ones) are claimed to exist for $\alpha $ close to 0.5.\cite{sushkov2001} Note also that the boundaries of the critical regions are not known exactly: for example, an $\alpha$ range from 0.34 to 0.60 is reported in Ref.~\onlinecite{sushkov2001} for the spin-liquid regime, while a wider region ($0.24-0.83$) is suggested in Ref.~\onlinecite{siurakshina2001}.

Despite numerous theoretical investigations, experimental realizations of
the $J_{1}-J_{2}$ model are scarce. Layered vanadium oxides Li$_{2}$VOXO$_{4}$ (X = Si, Ge) and VOMoO$_4$ were the first examples of the FSL systems. Initially, these materials were ascribed to $\alpha\simeq 1$ region of the phase diagram, and frustration-driven structural distortions were conjectured on the basis of nuclear magnetic resonance (NMR) data.\cite{melzi2000,melzi2001,carretta2002} However, later studies revealed the lack of the structural distortions and established a different scenario for all three systems. Thus, Li$_2$VOXO$_4$ compounds fall into the CAF region with $\alpha\gg 1$,\cite{rosner2002,rosner2003,bombardi2004} while VOMoO$_4$ lies in the NAF region with $\alpha\simeq 0.2$.\cite{bombardi2005}

Complex vanadium phosphates AA'VO(PO$_{4})_{2}$ present another realization of the FSL model with ferromagnetic $J_{1}$ and antiferromagnetic $J_{2}$ as will be discussed in detail below. We should
also mention two very recent propositions for the FSL systems. (CuX)LaNb$_{2} $O$_{7}$ (X = Cl, Br) perovskite-type compounds were claimed to realize the $J_{1}-J_{2}$ model with ferromagnetic $J_{1}$ and antiferromagnetic $J_{2}$.\cite{kageyama2005,kageyama2006} However, the estimates of $J$'s from different experimental techniques are contradictory, and the validity of the FSL model for these systems is still unclear.\cite{kageyama2007} The layered perovskite PbVO$_{3}$ also reveals an interesting square lattice system with $\alpha \sim 0.3$, i.e., quite close to the antiferromagnetic (AFM) critical region.\cite{tsirlin2008,azuma2008} However, the complicated preparation procedure strongly hampers detailed investigation of this compound. Thus, little experimental information about the critical regions of the FSL model is available, and the search for new FSL systems is still challenging.

\begin{figure}
\includegraphics[scale=1]{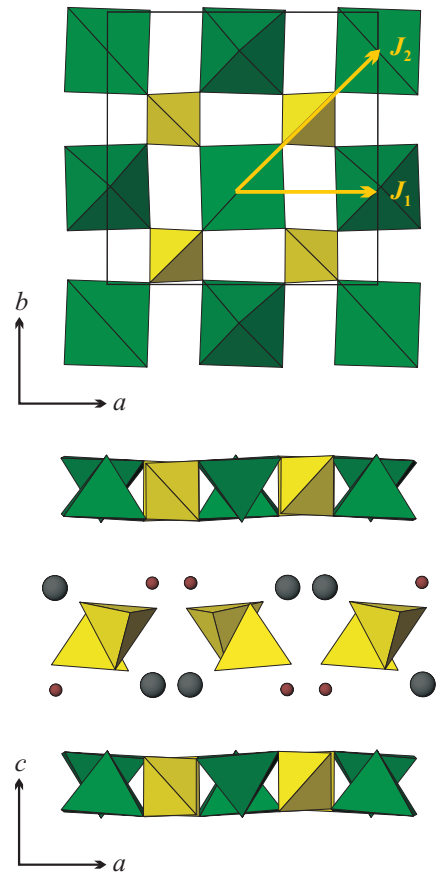}
\caption{\label{structure1}
(Color online) Crystal structure of BaCdVO(PO$_{4})_{2}$: single [VOPO$_{4}$] layer (upper panel) and stacking of the layers (bottom panel). Arrows indicate superexchange interactions $J_{1}$ (along the side of the square) and $J_{2}$ (along the diagonal of the square). Larger and smaller spheres denote Ba and Cd cations, respectively.}
\end{figure}

Complex vanadium phosphates AA'VO(PO$_{4})_{2}$ have layered crystal structures (Fig.~\ref{structure1}) with square lattice-like arrangement of V$^{+4}$ ($S=1/2$) cations. [VOPO$_{4}$] layers are formed by VO$_{5}$ square pyramids linked via PO$_{4}$ tetrahedra. The tetrahedra allow for superexchange interactions both along the side and along the diagonal of the square, hence the FSL-like spin system is formed. Metal cations and additional, isolated PO$_{4}$ tetrahedra are located between the layers. The layers are flexible towards buckling, therefore metal-oxygen distances can be tuned, and different metal cations can be accommodated within the structure. Magnetic properties of the compounds with AA' = Pb$_{2}$, SrZn, and BaZn have been recently investigated by means of thermodynamic measurements\cite{kaul2004,kaul2005} and neutron scattering.\cite{skoulatos2007,skoulatos2007a} We found that all the compounds fall to the CAF region of the FSL phase diagram. They reveal ferromagnetic $J_{1}$ and antiferromagnetic $J_{2}$ with $\alpha $ varying from $-1.8$ (AA' = Pb$_{2}$, BaZn) to $-1.1$ (AA' = SrZn). 

An advantage of studying vanadium phosphates is the low energy scale for the exchange couplings (below 10~K) that allows for high-field experiments providing additional information about the properties of the system.\cite{schmidt2008} Yet there is also one complication. The crystal structures of AA'VO(PO$_{4})_{2}$ do not have tetragonal symmetry; therefore, vanadium atoms do not form a regular square lattice. At first glance, the distortion of the square lattice is negligible. However, even a very slight alteration of the structure can lead to drastic changes in the exchange couplings as shown recently for Ag$_2$VOP$_2$O$_7$ (Ref.~\onlinecite{ag2vop2o7}). In case of the AA'VO(PO$_{4})_{2}$ compounds, we studied this issue in detail using band structure calculations and found considerable deviations from the square lattice model for some of the systems.\cite{tsirlin2009} Nevertheless, the deviation is really negligible for one of the compounds -- BaCdVO(PO$_4)_2$. Below, we present magnetic properties of this compound and interpret the results within the FSL model. A detailed examination of the appropriate spin models for other AA'VO(PO$_{4})_{2}$ compounds will be published elsewhere.\cite{tsirlin2009}

The crystal structure of BaCdVO(PO$_{4})_{2}$ has been reported by Meyer \textit{et al.},\cite{meyer1997} but magnetic properties of this compound were not investigated. In our work, we use a set of different thermodynamic measurements to study magnetic interactions in BaCdVO(PO$_{4})_{2}$. We show that BaCdVO(PO$_{4})_{2}$ is a FSL system with the frustration ratio $\alpha \approx -0.9$, i.e., lying very close the critical region of the FSL.

The paper is organized as follows: Sec.~\ref{experiments} deals with the experimental details. In Sec.~\ref{results}, we present our results on BaCd(VO)(PO$_{4}$)$_{2}$. Section~\ref{discussion} contains a detailed discussion of the results in the light of the $J_{1}-J_{2}$ model followed by our conclusions.

\section{Experimental details}

\label{experiments} Polycrystalline samples of BaCd(VO)(PO$_4)_2$ were
prepared by solid state reaction technique using BaCO$_3$, CdO, V$_2$O$_3$, V%
$_2$O$_5$, and (NH$_4)_2$HPO$_4$ as starting materials (all the chemicals
had at least 99.9\% purity grade). The process involved two steps. First,
the intermediate compound BaCdP$_2$O$_7$ was prepared by firing the
stoichiometric mixture of BaCO$_3$ and (NH$_4)_2$HPO$_4$ at $950$~$^{\circ}$%
C in air for 48 h with one intermediate grinding. In the second step,
stoichiometric amounts of BaCdP$_2$O$_7$, V$_2$O$_3$, and V$_2$O$_5$ were
grinded, pelletized and annealed in dynamic vacuum (10$^{-5}$ mbar) or
evacuated and sealed quartz tube (10$^{-2}$ mbar) at $800$ $^{\circ}$C for
30 h.

The phase composition of the prepared samples was checked by x-ray diffraction (XRD) (Huber G670f
camera, CuK$_{\alpha 1}$ radiation, ImagePlate detector). The samples
contained BaCd(VO)(PO$_{4})_{2}$ and minor amount \mbox{($\sim 3$\%)} of unreacted
BaCdP$_{2}$O$_{7}$. One then expects an equal amount of unreacted VO$_{2}$ to
be present in the samples. However, a minor impurity of VO$_{2}$ can not be
resolved by XRD due to the overlap of the strongest reflection of VO$_{2}$
with that of BaCdVO(PO$_{4})_{2}$. Instead, a minor amount of the VO$_{2}$
impurity is evidenced by a small kink in the magnetic susceptibility
data at $340$ K (see Sec.~\ref{results}).

We tried to improve the quality of the samples by varying temperature and duration of the annealing. Unfortunately, BaCdVO(PO$_{4})_{2}$ is rather unstable, and the formation of an unknown impurity phase was observed after long annealings at 800 $^{\circ }$C or any annealings above 800 $^{\circ }$C. The annealing at 900 $^{\circ }$C resulted in melting and complete decomposition (as seen by powder XRD) of the compound towards unknown phases.\cite{foot1} Regarding these difficulties, we did not attempt to grow single crystals of BaCdVO(PO$_4)_2$, because the data measured on polycrystalline samples are sufficient to determine the parameters of the FSL model and the location of the system within the phase diagram.

Magnetization ($M$) data were measured as a function of temperature using a
SQUID magnetometer (Quantum Design MPMS). Specific heat $C_{p}(T)$ was
measured on a pressed pellet with a standard relaxation technique using a
Quantum Design PPMS. All the measurements were carried out over a wide
temperature range ($0.4$ K $\leq $ $T$ $\leq $ $400$ K) and a field up to $7$
T. The low-temperature measurements were done partly using an additional $^{3}$He~setup.

\section{Results}

\label{results} 
\begin{figure}
\includegraphics[scale=1]{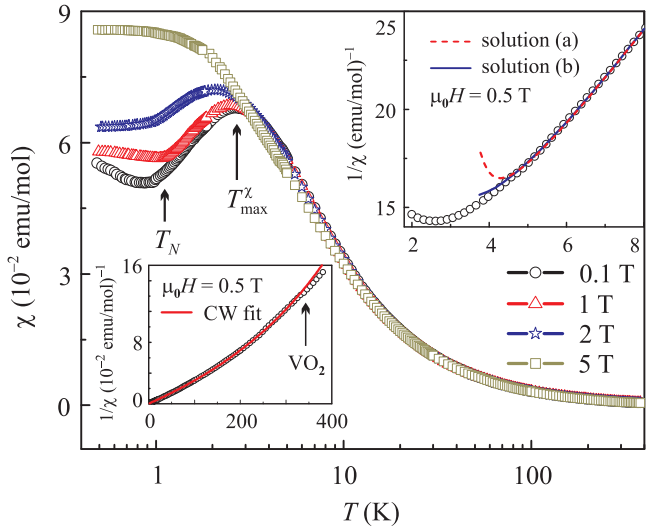}
\caption{\label{suscept1}
(Color online) Temperature dependence of the susceptibility $\chi(T)$ of BaCd(VO)(PO$_{4}$)$_{2}$ measured at different applied fields. Insets show inverse susceptibility (open circles) along with the HTSE [Eq. (\ref{HTSE}), upper inset] and the Curie-Weiss [Eq. (\ref{Curie}), lower inset] fits. In the upper inset, dashed and solid lines denote solutions (a) and (b), respectively. In the lower inset, solid line is the Curie-Weiss fit, while the arrow indicates the kink at 340 K due to the VO$_2$ impurity.}
\end{figure}

Figure~\ref{suscept1} shows the magnetic susceptibility $\chi =M/H$
of BaCdVO(PO$_{4})_{2}$ as a function of temperature. At high temperature
(above 20 K), the data behave in a Curie-Weiss (CW) manner. At lower
temperature and low fields, the curve exhibits a broad maximum at $T_{\max}^{\chi}\simeq 2.7$~K. The maximum is characteristic for low-dimensional spin systems and indicates a crossover to a state with antiferromagnetic correlations. A slight change of slope observed at $\sim 1$ K is associated
with magnetic ordering, while the Curie-like upturn in low fields below 1 K is
likely caused by extrinsic paramagnetic impurities. An increase of the magnetic
field leads to a shift of the maximum to lower temperatures, an increase
of the $\chi$ value at the maximum ($\chi_{\max }$), and an
increase of $\chi$ at the lowest investigated temperatures. This
behavior is quite similar to that observed in the other FSL systems\cite{kaul2005} and rather
typical for low-dimensional and/or frustrated antiferromagnetic systems. 

Careful examination reveals a small kink in $1/\chi $ vs. $T$ curve at $340$~K.
This kink is indicative of the VO$_{2}$ impurity, since VO$_{2}$ undergoes
metal-insulator transition at $340$~K.\cite{hyland1968} We emphasize that
this impurity does not affect any of the results reported below, as the
paramagnetic contribution of VO$_{2}$ below 300 K is
temperature-independent, while the energy scale of the exchange interactions
in BaCdVO(PO$_{4})_{2}$ is lower by two orders of magnitude. However, the
impurity prevents us from using the data above $300$~K in further analysis.

High-temperature ($20-300$ K) susceptibility data can be fitted with a 
CW law corrected for the temperature-independent contribution $\chi _{0}$ that
accounts for diamagnetism of core shells and Van Vleck paramagnetism: 
\begin{equation}
\chi (T)=\chi _{0}+\frac{C}{T+\theta_{\CW}}  \label{Curie}
\end{equation}%
The fitting resulted in $\chi _{0}=-4.4(1)\times 10^{-4}$ emu/mol, $C=0.375(2)$ emu~K/mol, $\theta_{\CW}=0.66(4)$~K. This $C$ value corresponds to an effective magnetic moment $\mu _{\text{eff}}=1.73(1)$ $\mu_{B}$ in perfect agreement with the expected spin-only value for V$^{+4}$. Note that $\theta_{\CW}$ is slightly dependent on the lower limit of the data used due to the curvature in $1/\chi$ related to the maximum in $\chi(T)$. However, values of $\theta_{\CW}$ below 1 K are obtained for any fit with a lower limit above 20 K.

Thus, we find a very low $\theta _{\CW}$ value, despite the susceptibility
maximum at $T_{\max }^{\chi }\simeq 2.7$ K indicates somewhat stronger exchange
interactions. This result implies that several interactions with different
signs are present in the system under investigation. To get a quantitative
estimate of the exchange interactions, we fit the susceptibility data above
5 K with the high-temperature series expansion (HTSE) for the FSL model:\cite{rosner2003}
\begin{equation}
\chi(T)=\chi_{0}+\frac{N_{A}g^{2}\mu_{B}^{2}}{k_{B}T}\sum_{n}\left(\frac{J_{1}}{k_{B}T}\right) ^{n}\sum_{m}c_{m,n}\left(\frac{J_{2}}{J_{1}}\right)^{m},
\label{HTSE}
\end{equation}
where $\chi_{0}$ is temperature-independent contribution, $g$ is Lande $g$-factor, and $c_{m,n}$ are the coefficients listed in Table I of Ref. \onlinecite{rosner2003}. We find $\chi _{0}=-3.8(2)\times 10^{-4}$ emu/mol, $g=1.968(3)$, $J_{1}=-3.62(5)$ K, and $J_{2}$ $=3.18(2)$ K [\textit{solution} (a)], or, alternatively, $\chi_{0}=-3.9(2)\times 10^{-4}$ emu/mol, $g=1.981(3)$, $J_{1}=2.16(1)$ K, and $J_{2}=-2.05(4)$ K [\textit{solution} (b)].\cite{foot4} The fit with the solution (b) looks somewhat better and extends to lower temperatures as compared to the fit with the solution (a) (see the upper inset of Fig.~\ref{suscept1}). Therefore, one may speculate that the solution (b) is the correct one. However, the difference between the two fits is caused by the lower exchange couplings in the solution (b), since the HTSE is valid at $T\geq J_i$. In the region used for the fitting (above 5~K), both the solutions produce the fits of similar quality, hence they are indistinguishable. The solution (a) locates the system in the CAF region close to the FM critical region (i.e., the region on the CAF--FM boundary), while the solution (b) corresponds to a position deep into the stable NAF region. The presence of two solutions in fitting $\chi(T)$ with the HTSE for the FSL is a well-known problem. The ambiguity has to be resolved by the use of other experimental data.

To discriminate the valid set of $J$'s, we turn to magnetization data and
analyze the value of the saturation field $H_{s}$. According to theoretical
results by Schmidt \textit{et al.},\cite{schmidt2007a} the saturation field can
be calculated as 
\begin{eqnarray}
\mu_0H_{s}&=&\frac{J_{c}k_{B}zS}{g\mu _{B}}\left[ \left( 1-\tfrac{1}{2}(\cos
Q_{x}+\cos Q_{y})\right)\right. \cos \varphi \nonumber\\
&&\left.+(1-\cos Q_{x}\cos Q_{y})\sin \varphi\right],
\end{eqnarray}
where $z=4$ (magnetic coordination number), $S=1/2$ is the spin value, $\varphi $ and $J_{c}$ are defined in Section~\ref{intro}, and $(Q_{x},Q_{y})$ is the wave vector of the ordered state. Using the appropriate wave vectors for the CAF and NAF regions, one finds $\mu_0H_{s}=2(J_{1}+2J_{2})k_B/(g\mu _{B})$ for CAF and $4J_{1}k_B/(g\mu _{B})$ for NAF phases. The first set of $J$'s [solution (a)] results in $\mu_0H_{s}=4.2$ T, while the second set [solution (b)] gives rise to a much higher saturation field $\mu_0H_{s}=6.55$ T.

\begin{figure}
\includegraphics[scale=1]{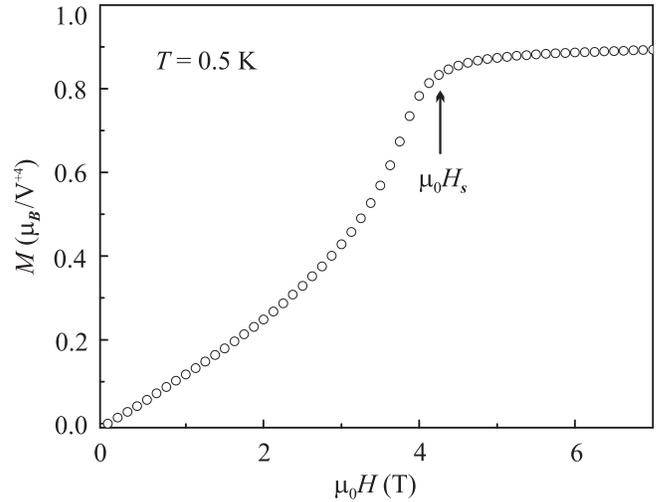}
\caption{\label{mvsh1}
Magnetization ($M$) as a function of the applied field ($H$) measured at $0.5$ K -- well below $T_{N}$. The arrow marks the saturation field $\mu_0H_s$.}
\end{figure}

Figure~\ref{mvsh1} presents the magnetization curve measured at 0.5 K, i.e., well
below the ordering temperature $T_{N}$. At low fields, the $M(H)$ dependence is
linear, while a positive curvature is observed above 2 T leading to a marked
kink at the saturation field $\mu_0H_{s}\simeq 4.2$ T.\cite{foot2} The saturation magnetization 
$M_s\simeq 0.9\ \mu_{B}$/V$^{+4}$ is slightly less than the expected value of $1\ \mu_{B}$. The reduction of $M_s$ is likely caused by the weight error due to the presence of non-magnetic impurity (BaCdP$_2$O$_7$) in the samples under investigation (see Sec.~\ref{experiments}).\cite{foot5} The curvature above 2 T is also somewhat puzzling. At first glance, one may interpret it as a spin-flop transition. However, the magnetic anisotropy of V$^{+4}$ is weak. Thus, a spin-flop transition in Pb$_2$VO(PO$_4)_2$ is observed at 1 T,\cite{kaul2005} and a strong increase of the anisotropy in the isostructural BaCdVO(PO$_4)_2$ compound is unlikely. In the next section, we will suggest an alternative and plausible explanation for the curvature of the $M(H)$ dependence above 2 T.

The experimentally determined saturation field matches exactly the
value calculated for the solution (a) of the susceptibility fit but is far
below the value expected for the solution (b). This clearly demonstrates that
the solution (a), with a ferromagnetic $J_{1}$ and an antiferromagnetic $%
J_{2}$ is the correct one. Thus, BaCdVO(PO$_{4}$)$_{2}$ belongs to the
frustrated ferromagnetic square lattice system as all the other AA$^{^{\prime }}$%
VO(PO$_{4}$)$_{2}$ structural homologs.

\begin{figure}
\includegraphics[scale=1]{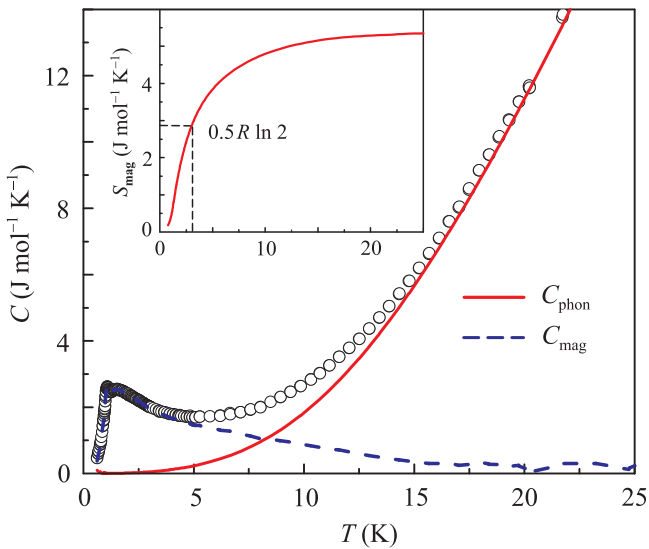}
\caption{\label{cp1}
(Color online) Temperature dependence of the specific heat $C_{p}(T)$ of BaCd(VO)(PO$_{4}$)$_{2}$ measured at zero applied field. Open circles are the raw data, the solid line shows the phonon contribution [according to the fit with Eq. (\ref{Debye})], and the dashed line indicates the magnetic contribution $C_{\mg}$. The anomaly at $T_N\simeq 1$~K is shown in detail in Fig.~\ref{cph1}. In the inset, the magnetic entropy ($S_{\mg}$) is plotted as a function of temperature, dashed lines show the entropy of $0.5R\ln 2$ and the respective temperature.}
\end{figure}

\begin{figure}
\includegraphics[scale=1]{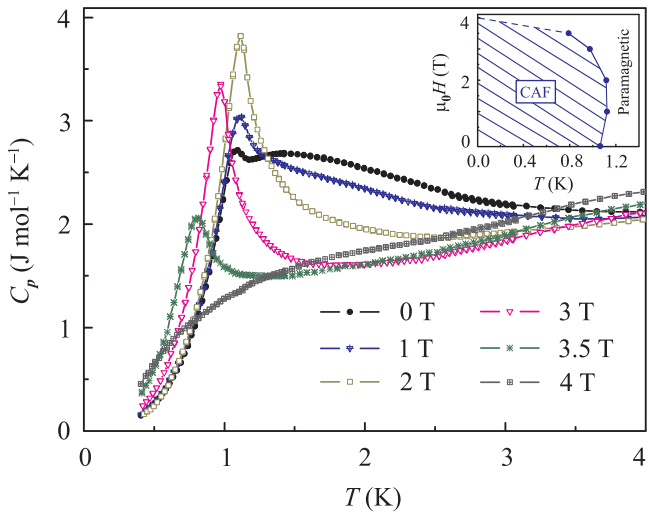}
\caption{\label{cph1}
Specific heat of BaCdVO(PO$_{4})_{2}$ measured in different magnetic fields. The inset shows the $H-T$ phase diagram of the system.}
\end{figure}

Specific heat ($C_{p}$) measurement at zero field is shown in Fig.~\ref{cp1}. At high temperatures, $C_{p}$ is completely dominated by phonon excitations. Below 5 K, a decrease of temperature is accompanied by an increase of $C_{p}$ indicating that the magnetic contribution to the specific heat becomes prominent. At low temperatures, $C_{p}(T)$ shows a broad maximum at $T_{\max}^{\,C}\simeq 1.5$~K due to the spin correlations and a small but well resolved peak at $T_{N}\simeq 1$ K associated with the magnetic ordering (see also Fig.~\ref{cph1}). Below $T_{N}$, $C_{p}(T)$ drops rapidly.

To extract the magnetic contribution to the specific heat $C_{\mg}$, we
subtract an estimated phonon contribution from the total measured specific
heat $C_{p}$. For this purpose, the experimental data at high temperatures ($%
15$~K $\leq T\leq 200$~K) were fitted with 
\begin{equation}
C_{p}(T)=\frac{A}{T^{2}}+9R\sum_{n=1}^{n=4}c_{n}\left( \frac{T}{\theta_D^{(n)}}\right)^{3}\int\limits_{0}^{\theta_D^{(n)}/T}\frac{x^{4}e^{x}}{\left( e^{x}-1\right)^2}\,dx,
\label{Debye}
\end{equation}%
where $A/T^{2}$ accounts for the magnetic contribution at high temperatures, 
$R=8.314$ \mbox{J/mol K} is the gas constant, and the sum of Debye
functions accounts for the phonon contribution. The use of several Debye
functions with distinct characteristic temperatures $\theta_D^{(n)}$ is
necessary due to the large difference in atomic masses of the elements
forming the BaCdVO(PO$_{4})_{2}$ compound. In general, the procedure is similar to that
reported in Refs.~\onlinecite{kaul2005}, \onlinecite{kini2006}, and \onlinecite{nath2008} for the related vanadium phosphates. The resulting $C_{\mg}(T)$ curve is shown in Fig.~\ref{cp1}. In the
inset, we plot the temperature dependence of the magnetic entropy $S_{\mg}(T)$
as obtained by integrating $C_{\mg}(T)/T$. At high temperatures, $S_{\mg}(T)$
converges towards the value $S_{\infty }\simeq 5.3$ J/mol~K, in reasonable
agreement with the expected value $R\ln 2=5.76$ J/mol~K taking into account
the uncertainty in the estimated phonon contribution. The result indicates that
$C_{\mg}$ reflects the intrinsic contribution of BaCdVO(PO$_{4})_{2}$. 

At the lowest limit ($T=15$ K) of the specific heat fit to Eq. (\ref{Debye}), the magnetic contribution $A/T^{2}$ amounts to a small part of the total specific heat only. Therefore, the parameter $A$ is obtained with a large uncertainty, $A=100\pm 30$ J K/mol. The $A/T^2$ term corresponds to the lowest order in the HTSE for the specific heat\cite{rosner2003} and may be expressed as $A=0.375J_{c}=0.375R(J_{1}^{2}+J_{2}^{2})$. We find $A=72$ \mbox{J K/mol} and 
$A=29$ \mbox{J K/mol} using the $J_i$ values of the solutions (a) and (b) of the
susceptibility fit, respectively. Both theoretical values are below the experimental one, but the result for the solution (a) is still within the error bar, while that for the solution (b) is far below. Thus, this analysis also favors the solution (a) and the CAF scenario for BaCdVO(PO$_4)_2$.

A further rough estimate of the thermodynamic energy scale $J_{c}$ can be
obtained from the $T$-dependence of the magnetic entropy. In a general approximation, 
$J_{c}$ is approximately twice the temperature at which the entropy reaches
half of its high-temperature limit. In BaCdVO(PO$_{4})_{2}$, $S(T)=0.5R\ln 2$
at $T=2.5$ K. This temperature corresponds to $J_{c}=5$~K, in perfect agreement
with the solution (a) ($J_{c}=4.8$ K), but in poor compliance with the solution (b) ($J_{c}=3.0$ K).

Field dependent specific heat measurements are presented in Fig.~\ref{cph1}.
The increase of the field results in the suppression of the broad maximum at $1.5$ K,
while the absolute value of $C_{p}$ at $T_{N}$ is enhanced. The maximum
enhancement of the peak value is observed at $2$ T where the transition
anomaly is most pronounced. With further increase of the field, the peak value
decreases again and the anomaly broadens slightly. However, between $3.5$ and $4$ T the anomaly gets completely suppressed. The variation of the transition temperature is as follows: $T_{N}$ increases very slightly from 0 to 2 T up to a maximum $T_{N}(2\text{ T})=1.1$~K and decreases more
clearly with further increase of the field. This result enables to draw the $H-T$ phase diagram of the system (inset of Fig.~\ref{cph1}). The field dependence of the specific heat is similar to that of Pb$_{2}$VO(PO$_{4})_{2}$,\cite{kaul2005} but the smaller energy scale of the exchange couplings
facilitates experimental access to the larger part of the $H$ vs. $T$ phase diagram. 

The field dependence may be understood as follows. The broad maximum at $1.5$ K is
caused by short-range antiferromagnetic correlations. These correlations are
suppressed by the applied magnetic field, hence the maximum is strongly
reduced already at $1$ T. The magnetic entropy is transferred to the
transition anomaly resulting in a much larger and sharper peak at $T_{N}$. As
we have mentioned in the introduction, long-range magnetic ordering in
low-dimensional and/or frustrated spin systems is suppressed by quantum
fluctuations. Magnetic field suppresses these fluctuations, therefore $T_{N}$
is slightly enhanced at low fields. However, above 2 T the field is strong
enough to overcome the antiferromagnetic ordering, hence $T_{N}$ is reduced and
finally suppressed below 0.4~K at $4$~T.

\section{Discussion}

\label{discussion}Our experimental results and our analysis demonstrate that
BaCdVO(PO$_{4}$)$_{2}$ is a frustrated square lattice with a ferromagnetic
exchange $J_{1}$ along the side of the square and an antiferromagmetic
exchange $J_{2}$ along the diagonal. The first evidence for the presence of
antiferromagnetic and ferromagnetic exchange of similar size is given by the
very small value of the Weiss constant \mbox{$\theta _{\CW}$ $<1$~K} contrasting a
much higher value of the temperature $T_{\max}^{\chi}\simeq 2.7$~K of the maximum
in $\chi (T)$. The latter one reflects the onset of antiferromagnetic
correlations. In a non-frustrated $S=1/2$ antiferromagnetic square lattice,
this maximum is expected at $T_{\max}^{\chi}=0.95J$ (Refs. \onlinecite{makivic1991} and \onlinecite{troyer1998}), suggesting some antiferromagnetic exchange with $J\gtrsim 3$ K in BaCdVO(PO$_{4}$)$_{2}$. Then a much smaller $\theta_{\CW}$, which is the sum of all the exchange interactions in the system under investigation, implies an additional exchange of the opposite sign and thus ferromagnetic.

A fit of the susceptibility in the $T$ range $5$ K $\leq $ $T$ $\leq $ $300$ K with the HTSE for the FSL resulted in two sets of exchange coupling parameters: solution (a) with $J_{1}\simeq -3.6$~K, $J_{2}\simeq 3.2$~K and solution (b) with $J_{1}\simeq 2.2$~K, $J_{2}\simeq -2.1$~K. The two solutions correspond to different regions of the FSL phase diagram and realize the frustrated CAF and non-frustrated NAF regimes, respectively. Further experimental results enable to select (a) as the correct solution. 

The frustrated regime of the solution (a) causes the low saturation field of $\mu_0H_s=4.2$ T. The non-frustrated ground state for the solution (b) corresponds to the higher saturation field of $\mu_0H_s=6.55$ T despite both $J_1$ and $J_2$ are weaker than in the solution (a). Magnetization measurements at low temperatures show a well defined saturation at the saturation field $\mu_0H_{s}=4.2$ T. This value matches perfectly the $\mu_0H_{s}$ expected for the solution (a) but is well below the $\mu_0H_{s}$ expected for the solution (b). Further on, an analysis of the magnetic specific heat at high temperatures ($>15$ K) as well as of the $T$ dependence of the magnetic entropy at low temperatures ($<3$ K) leads to a thermodynamic energy scale of the exchange couplings $J_{c}$ in reasonable agreement with the solution (a) but in clear disagreement with the solution (b). Therefore, the solution (a) is the appropriate one to describe the spin system of BaCdVO(PO$_{4}$)$_{2}$.

The solution (a) with $J_{1}\simeq -3.6$ K and $J_{2}\simeq 3.2$ K places the system to the CAF region of the FSL phase diagram. With $J_{2}/J_{1}=-0.9$ corresponding to $\varphi/\pi =0.76$, BaCdVO(PO$_{4}$)$_{2}$ is quite close to the border of the FM critical regime ($J_{2}/J_{1}=-0.7$), closer than any of the previously reported FSL compounds such as SrZnPO(PO$_{4}$)$_{2}$ ($J_{2}/J_{1}=-1.1$) and Pb$_{2}$VO(PO$_{4}$)$_{2}$ or BaZnVO(PO$_{4}$)$_{2}$ ($J_{2}/J_{1}=-1.8$). A direct evidence for the strong frustration in BaCdVO(PO$_{4}$)$_{2}$ is given by the position of the specific heat maximum ($T_{\max}^{\,C}$) and the maximum value of the magnetic specific heat ($C_{\max}$). The specific heat of the FSL at low temperatures ($T<J_{c}$) is not known precisely, since there is presently no established way to compute it. However, for an unfrustrated square lattice, the value at the maximum seems nowadays to be established within a few percent, $C_{\max}\simeq 0.46R=3.82$ J/mol~K at $T_{\max}^{\,C}/J_{1}=0.60$.\cite{makivic1991,hofmann2003} It is further established that tuning the FSL towards a critical region leads to a broadening of this maximum, a reduction of the $C_{\mg}$ value at the maximum, and a shift of the maximum towards lower temperatures. Such a trend was already observed in the previously studied FSL systems, where $C_{\max}/R$ and $T_{\max}^{\,C}/J_{c}$ are decreased from $0.46$ and $0.6$ in Li$_2$VOSiO$_{4}$ (far away from any critical region) towards $0.44$ and $0.40$ in Pb$_{2}$VO(PO$_{4}$)$_{2}$ and finally to $0.40$ and $0.29$ in SrZnVO(PO$_{4}$)$_{2}$ (closer to the critical region).\cite{kaul2005} 

In BaCdVO(PO$_{4}$)$_{2}$, $C_{\max}/R$ reaches only $0.33$, well below the value found in the other FSL systems, while $T_{\max}^{\,C}/J_{c}=0.31$. The very small value of $C_{\max}$ cannot be attributed to a scaling problem due to a large amount of a foreign phase, since the magnetic entropy at high temperatures is close to $R\ln 2$ (the phonon contribution at $T_{\max}^{\,C}$ amounts to less than $1$\% of the total specific heat and can therefore safely be neglected). Further on, a comparison of $C_{\mg}(T)$ near its maximum for different FSL systems using a reduced $T/J_{c}$ temperature scale indicates that the maximum in $C_{\mg}(T)$ of BaCdVO(PO$_{4}$)$_{2}$ is broader, and the decrease of $C_{\mg}(T)$ above $T_{\max}^{\,C}$ is much weaker than in the other known FSL systems. Thus, this broad maximum with a small $C_{\mg}$ value at the maximum is also a direct experimental evidence for the strong frustration.

A further (and new) manifestation of the frustration is likely observed in our magnetization curve (Fig. \ref{mvsh1}). Above $2$ T, the slope of $M(H)$ increases with magnetic field and steeps up just before saturation is reached. Such a behavior has been recently predicted by theoretical calculations for the FSL.\cite{schmidt2008} This steeping up just below $H_s$ is suggested to be more pronounced for systems that are close to the critical regions. However, further magnetization measurements on other FSL systems are needed to confirm this trend.

Using exact diagonalization of finite-size clusters, Shannon \textit{et al.}\cite{shannon2004} calculated the dependence of a number of characteristic properties of the FSL (such as $T_{\max}^{\chi}$, $T_{\max}^{\,C}$, $\chi_{\max}$, and $C_{\max}$) as a function of $\varphi$. Comparison of these predictions with the experimental observation for BaCdVO(PO$%
_{4}$)$_{2}$ and the related compounds indicates that while the overall $\varphi$ dependence seems to be reproduced, the calculated absolute values are slightly too large. This difference is likely a consequence of the finite cluster size that leads to a gapped energy excitation spectra and thus an exponential vanishing of $C_{\mg}(T)$ and $\chi (T)$ at very low $T$ in contrast to the real behavior. The missing entropy has then to be recovered at higher temperatures, thus enhancing $C_{\mg}(T)$ at its maximum and shifting the maximum to higher $T$.

A last comment on the antiferromagnetic order observed at $T_{N}\simeq 1.0$~K. The anomalies in $\chi (T)$ and $C_{p}(T)$ are much smaller than those expected for a classical three-dimensional magnetic system but very similar to those observed in the other FSL systems, e.g., Li$_{2}$VOSiO$_{4}$ (Refs.~\onlinecite{melzi2001} and \onlinecite{kaul2005}) or Pb$_{2}$VO(PO$_{4}$)$_{2}$ (Refs.~\onlinecite{kaul2004} and \onlinecite{kaul2005}). Since in some of these systems long-range magnetic order was directly confirmed by NMR or neutron scattering experiments,\cite{bombardi2004,skoulatos2007,skoulatos2007a} the similarity in the behavior allows us to safely claim that these small anomalies in $\chi(T)$ and $C_p(T)$ correspond to the onset of (columnar) antiferromagnetic order in BaCdVO(PO$_{4}$)$_{2}$ too. The ratio $R=T_{N}/T_{\max}^{\chi}$ is smaller in BaCdVO(PO$_{4}$)$_{2}$ ($R=0.39$) than in the other FSL systems ($0.4<R<0.6$), reflecting either the weakness of interlayer exchange and/or the suppression of the AFM order by quantum fluctuations.

In conclusion, our study of BaCdVO(PO$_{4})_{2}$ shows that the magnetic properties of this compound are well understood within the FSL model. Magnetic susceptibility, magnetization, and specific heat data consistently suggest $J_{1}\simeq -3.6$ K, $J_{2}\simeq 3.2$ K, hence $\alpha \simeq -0.9$, locating the compound into the CAF region of the FSL phase diagram. BaCdVO(PO$_{4})_{2}$ lies closer to the critical region of the FSL than any of the previously reported compounds. This conclusion is supported by a strongly reduced maximum of the magnetic specific heat and a positive curvature of the magnetization curve consistent with the recent theoretical predictions. 

\begin{acknowledgments}
The authors are grateful to Nic Shannon and B. Schmidt for fruitful discussions. Financial support of GIF (Grant No. I-811-257.14/03), RFBR (Grant No. 07-03-00890), and the Emmy Noether program of the DFG is acknowledged.
\end{acknowledgments}


\begin{thebibliography}{99}
\bibitem{anderson1987} P. W. Anderson, Science \textbf{235}, 1196 (1987).

\bibitem{lee2008} P. A. Lee, Rep. Prog. Phys. \textbf{71}, 012501 (2008).

\bibitem{shannon2004} N. Shannon, B. Schmidt, K. Penc, and P. Thalmeier,
Eur. Phys. J. B \textbf{38}, 599 (2004); cond-mat/0312160.

\bibitem{chandra1988} P. Chandra and B. Doucot, Phys. Rev. B \textbf{38},
9335 (1988).

\bibitem{sushkov2001} O. P. Sushkov, J. Oitmaa, and Z. Weihong, Phys. Rev. B 
\textbf{63}, 104420 (2001); cond-mat/0007329, and references therein.

\bibitem{siurakshina2001} L. Siurakshina, D. Ihle, R. Hayn, Phys. Rev. B 
\textbf{64}, 104406 (2001), and references therein.

\bibitem{bacci1991} S. Bacci, E. Gagliano, and E. Dagotto, Phys. Rev. B 
\textbf{44}, 285 (1991).

\bibitem{shannon2006} N. Shannon, T. Momoi, and P. Sindzingre, Phys. Rev.
Lett. \textbf{96}, 027213 (2006); cond-mat/0512349.

\bibitem{schmidt2007a} B. Schmidt, P. Thalmeier, and N. Shannon, Phys. Rev.
B \textbf{76}, 125113 (2007); arXiv:0705.3094.

\bibitem{schmidt2007b} B. Schmidt, N. Shannon, and P. Thalmeier, J. Phys.:
Condens. Matter \textbf{19}, 145211 (2007).

\bibitem{schmidt2008} P. Thalmeier, M. E. Zhitomirsky, B. Schmidt, and N.
Shannon, Phys. Rev. B \textbf{77}, 104441 (2008); arXiv:0711.4054.

\bibitem{foot3}Note that this phase is commonly named "collinear antiferromagnet", although the term is somewhat misleading.

\bibitem{melzi2000} R. Melzi, P. Carretta, A. Lascialfari, M. Mambrini, M.
Troyer, P. Millet, and F. Mila, Phys. Rev. Lett. \textbf{85}, 1318 (2000); cond-mat/0005273.

\bibitem{melzi2001} R. Melzi, S. Aldrovandi, F. Tedoldi, P. Carretta, P. Millet, and F. Mila, Phys. Rev. B \textbf{64}, 024409 (2001); cond-mat/0101066.

\bibitem{carretta2002} P. Carretta, N. Papinutto, C. B. Azzoni, M. C. Mozzati, E. Pavarini, S. Gonthier, and P. Millet, Phys. Rev. B \textbf{66}, 094420 (2002).

\bibitem{rosner2002} H. Rosner, R. R. P. Singh, W. H. Zheng, J. Oitmaa,
S.-L. Drechsler, and W. E. Pickett, Phys. Rev. Lett. \textbf{88}, 186405
(2002); cond-mat/0110003.

\bibitem{rosner2003} H. Rosner, R. R. P. Singh, W. H. Zheng, J. Oitmaa, and
W. E. Pickett, Phys. Rev. B \textbf{67}, 014416 (2003).

\bibitem{bombardi2004} A. Bombardi, J. Rodriguez-Carvajal, S. Di Matteo, F. de Bergevin, L. Paolasini, P. Carretta, P. Millet, and R. Caciuffo, Phys. Rev. Lett. \textbf{93},
027202 (2004).

\bibitem{bombardi2005} A. Bombardi, L. C. Chapon, I. Margiolaki, C. Mazzoli, S. Gonthier, F. Duc, and P. G. Radaelli, Phys. Rev. B \textbf{71}, 220406(R) (2005).

\bibitem{kageyama2005} H. Kageyama, T. Kitano, N. Oba, M. Nishi, S. Nagai,
K. Hirota, L. Viciu, J. B. Wiley, J. Yasuda, Y. Baba, Y. Ajiro, and K.
Yoshimura, J. Phys. Soc. Jpn. \textbf{74}, 1702 (2005).

\bibitem{kageyama2006} N. Oba, H. Kageyama, T. Kitano, J. Yasuda, Y. Baba,
M. Nishi, K. Hirota, Y. Narumi, M. Hagiwara, K. Kindo, T. Saito, T. Ajiro,
and K. Yoshimura, J. Phys. Soc. Jpn. \textbf{75}, 113601 (2006).

\bibitem{kageyama2007} M. Yoshida, N. Ogata, M. Takigawa, J. Yamaura, M.
Ichihara, T. Kitano, H. Kageyama, Y. Ajiro, and K. Yoshimura, J. Phys. Soc.
Jpn. \textbf{76}, 104703 (2007); arXiv:0706.3559.

\bibitem{tsirlin2008} A. A. Tsirlin, A. A. Belik, R. V. Shpanchenko, E. V.
Antipov, E. Takayama-Muromachi, and H. Rosner, Phys. Rev. B \textbf{77}, 092402 (2008); arXiv:0801.1434.

\bibitem{azuma2008} K. Oka, I. Yamada, M. Azuma, S. Takeshita, K. H. Satoh, A. Koda, R. Kadono, M. Takano, and Y. Shimakawa, Inorg. Chem. \textbf{47}, 7355 (2008).

\bibitem{kaul2004} E. E. Kaul, H. Rosner, N. Shannon, R. V. Shpanchenko, and
C. Geibel, J. Magn. Magn. Mater. \textbf{272-276}, 922 (2004).

\bibitem{kaul2005} E. E. Kaul, PhD thesis, Technical University Dresden, Dresden (2005). Electronic version available at: {http://hsss.slub-dresden.de/documents/1131439690937-4924/1131439690937-4924.pdf}

\bibitem{skoulatos2007} M. Skoulatos, J. P. Goff, N. Shannon, E. E. Kaul, C. Geibel, A. P. Murani, M. Enderle, and A. R. Wildes, J. Magn. Magn. Mater. \textbf{310}, 1257 (2007).

\bibitem{skoulatos2007a} M. Skoulatos. Abstracts of the 4th Europian conference on Neutron scattering in Lund, Sweden, p. 164 (2007).

\bibitem{ag2vop2o7}A. A. Tsirlin, R. Nath, C. Geibel, and H. Rosner, Phys. Rev. B \textbf{77}, 104436 (2008); arXiv:0802.2293.

\bibitem{tsirlin2009}A. A. Tsirlin and H. Rosner, unpublished results on the band structures of the AA'VO(PO$_4)_2$ compounds. 

\bibitem{meyer1997} S. Meyer, B. Mertens, and Hk. M{\"u}ller-Buschbaum, Z.
Naturforsch. \textbf{52b}, 985 (1997).

\bibitem{foot1} Our results are somewhat different from that of Meyer \textit{et al.}\cite{meyer1997} who report the preparation of single crystals at 975~$^{\circ}$C (i.e., above the decomposition temperature of 900~$^{\circ}$C observed in our experiments). The origin of this discrepancy is not yet clear. We should emphasize that Meyer \textit{et al.}\cite{meyer1997} report
the preparation of small single crystals only. No information about the synthesis and thermal properties of bulk samples of BaCdVO(PO$_4)_2$ is available.

\bibitem{hyland1968} G. J. Hyland, J. Phys. C \textbf{1}, 189 (1968).

\bibitem{foot4}The standard deviations for the exchange integrals originate from the fitting procedure only. Varying the lower limit of the data used, one finds a more pronounced change of $J$'s, and the actual error bar is of the order of 0.1 K.

\bibitem{foot2}We define the saturation field $H_s$ as a point of the maximum negative curvature of the $M(H)$ curve.

\bibitem{foot5}Note that we observe a similar {($\sim 7$\%)} reduction of the magnetic entropy in our specific heat data.

\bibitem{kini2006} N. S. Kini, E. E. Kaul, and C. Geibel, J. Phys.: Condens.
Matter, \textbf{18}, 1303 (2006).

\bibitem{nath2008} R. Nath, A. A. Tsirlin, E. E. Kaul, M. Baenitz, N. B\"uttgen, C. Geibel, and H. Rosner, Phys. Rev. B \textbf{78}, 024418 (2008), arXiv:0804.4667.

\bibitem{makivic1991}M. S. Makivi\'c and H.-Q. Ding, Phys. Rev. B \textbf{43}, 3562 (1991).

\bibitem{troyer1998}J.-K. Kim and M. Troyer, Phys. Rev. Lett. \textbf{80}, 2705 (1998).

\bibitem {hofmann2003}M. Hofmann, T. Lorenz, K. Berggold, M. Gr\"uninger, A.
Freimuth, G. S. Uhrig, and E. Br\"uck, Phys. Rev. B \textbf{67}, 184502 (2003).
\end{thebibliography}
\end{document}